\documentclass[final,3p,times,twocolumn]{elsarticle}

\usepackage{amssymb}

\usepackage{eufrak}
\newcommand{\goth}[1]{\mathfrak{#1}}
\journal{Physica {\bf B}}

\begin{document}

\begin{frontmatter}



\title{
\begin{minipage}[t]{7.0in}
\scriptsize
\begin{quote}
\leftline{{\it Physica B,} in press}
\end{quote}
\end{minipage}
\medskip
A Description of Phases with Induced Hybridisation at Finite Temperatures}


\author{D. I. Golosov}
\ead{Denis.Golosov@biu.ac.il}
\address{Department of Physics and the Resnick Institute, Bar-Ilan 
University, Ramat-Gan 52900, Israel.}

\begin{abstract}
In an extended Falicov-Kimball model, an excitonic insulator phase can be
stabilised at zero temperature. With increasing temperature, the excitonic
order parameter (interaction-induced hybridisation on-site, characterised by the
absolute value and phase) eventually becomes disordered, which involves 
fluctuations of both its phase and (at higher T) its absolute value. In order 
to build an adequate mean field description, it is important to clarify the 
nature of degrees of freedom associated with the phase and absolute value of 
the induced hybridisation, and the corresponding phase space volume. 
We show that a possible description is 
provided by the SU(4) parametrisation on-site. In principle, this allows to 
describe both the lower-temperature regime where phase fluctuations destroy 
the long-range order, and the higher temperature crossover corresponding to a 
decrease of absolute value of the hybridisation relative to the fluctuations 
level. This picture is also expected to be relevant in 
other contexts, including the Kondo lattice model. 
\end{abstract}


\end{frontmatter}


\section{Introduction}
\label{sec:intro}

The notion of induced hybridisation is familiar in many different contexts, 
including excitonic insulators\cite{Kohn,Khomskii76,Leder78}, 
Kondo insulators\cite{Mott74}, and superconductors 
(where a somewhat similar role can be played by the pairing 
amplitude\cite{NSR}). When the 
underlying non-interacting system is characterised by several different energy 
scales, the resultant behaviour at finite temperatures may prove rich and 
complex, as illustrated by the extended Falicov--Kimball model (FKM)
\cite{Zlatic},
\begin{eqnarray}
{\cal H}=&-&\frac{t}{2}\sum_{\langle i j \rangle} \left(c^\dagger_i c_j +
c^\dagger_j c_i \right) + E_d \sum_i d^\dagger_i d_i + \nonumber \\
&+&U \sum_i c^\dagger_i
d^\dagger_i d_i c_i + \delta{\cal H}\,,
\label{eq:FKM}
\end{eqnarray}
Here, the fermionic operators $c_i$ and
$d_i$ annihilate spinless fermions in the itinerant and (nearly) localised
band (the former with nearest-neighbour hopping amplitude $t$, the latter with 
the bare energy $E_d$), and $U$ is the strength of on-site repulsion 
(of the order of itinerant bandwidth or smaller). $\delta{\cal H}$ is a weak 
perturbation
(characteristic energy scale much less than $t$), which breaks the continuous
local degeneracy of the pure FKM with respect to the phases of operators $d_i$ 
[{\it i.e.,} $d_i \rightarrow \exp(i\phi_i) d_i$]. This could be exemplified 
by a weak nearest-neighbour hopping in the $d$-band, $\delta {\cal H}=- 
(t^{\,\prime}/2)\sum_{\langle i j \rangle} (d^\dagger_i d_j +
d^\dagger_j d_i ) $. 

Extensive investigations of the half-filled ($n=1$) case 
showed\cite{Czycholl99,Batista02,Farkasovsky08,prb12} that at 
$T=0$ a sizeable region of parameter 
space exists, whereby the ground state of the system is an excitonic 
condensate, or equivalently an excitonic insulator with long-range order. 
The order parameter is the induced hybridisation\cite{Khomskii76,Leder78},
\begin{equation}
\Delta_i=\langle c^\dagger_i d_i \rangle
\end{equation}
(or a Fourier harmonic of it), which in principle can reach the order of unity. 
For simplicity, here we will speak about the case of a uniform 
$\Delta_i \equiv \Delta$. Within the Hartree--Fock mean-field description, 
$\Delta$ solves a BCS-type equation. However, in a marked difference 
from the BCS scenario, the zero-temperature hybridisation gap $2U\Delta$ does 
not determine the scale of critical temperature $T_c$ beyond which the 
long-range order is lost. Instead, the scale\footnote{Here and below, 
temperature is measured in energy 
units, 
setting $k_B=1$.} 
of $T_c \ll 2 U \Delta$ is that  
of the low-lying collective excitations\cite{prb12} at $T=0$, which in turn is 
dictated 
by $\delta {\cal H}$. 

Indeed, at $\delta {\cal H}=0$ in an (unstable) $\Delta_i \equiv \Delta$ 
state there exists an entire
excitation branch with identically vanishing energy, as a consequence of 
continuous local degeneracy. At $T=0$, excitonic insulator state is stabilised 
once this
branch acquires positive energy at all momenta (except possibly for 
isolated Goldstone modes), 
which requires a parametrically small but finite 
perturbation\cite{prb12} ({\it e.g.,} 
$t^{\,\prime} < t^{\,\prime}_{cr}$ with $0< -t^{\,\prime}_{cr} \ll t$). The value of $T_c$ is
then determined by the characteristic 
energy of this low-lying branch [{\it e.g.,} roughly 
$\propto \sqrt{|t^{\,\prime}|(|t^{\,\prime}|-|t^{\,\prime}_{cr}|)}$]. Since the degeneracy at 
$\delta {\cal H}=0$ is associated with the phases of $d_i$, or equivalently with
those of $\Delta_i$, it is clear that the low-lying excitations at small
$\delta {\cal H}$ correspond to deviations of {\it phases} (as opposed to the 
amplitudes) of
$\Delta_i$ from the uniform constant value, and the transition at $T_c$ 
corresponds to a loss of long-range order of these phases. Breaking 
the individual electron-hole pairs, on the
contrary, requires a much larger energy of the order of  $2U|\Delta|$.

Above the second-order transition at $T=T_c$, the {\it phases} of $\Delta_i$ 
become disordered\cite{prb12,Apinyan}, whereas the fluctuations of the 
amplitude are still weak,
\begin{equation}
\Delta_i = |\Delta_i| \exp(i\phi_i)\,,\,\,\,\, |\Delta_i| \approx \Delta(T)\,.  
\end{equation}
While $\Delta(T)$ differs from zero, it is no longer associated with a 
symmetry breaking.
We note the similarity to the Kondo insulator\cite{Piers}, or to the pre-formed 
pairs above a superconducting transition\cite{Randeria,Kathy}.

It appears that the available mean-field results (see, {\it e.g.,}, 
Refs. \cite{Apinyan,Phan,Schneider08})  
lend support to a generic 
intuitive expectation that $\Delta(T)$ decreases via a smooth 
crossover\footnote{Reported phase transition at $T_\Delta$ is an artefact of
the methods used in Ref. \cite{Schneider08}.} at a 
temperature $T_\Delta$, which is roughly of the order of the zero-temperature 
hybridisation gap, 
$T_\Delta \sim 2 U\Delta(0)$. Beyond $T_\Delta$, the value 
of $\Delta(T)=\langle |\Delta_i| \rangle $ is comparable to the 
fluctuations of $ |\Delta_i|$.

In the case of the FKM, the crucial variables 
(such as the hybridisation amplitude $\Delta_i$) are defined on-site, which
suggests that a single-site mean-field theory might prove a useful starting
point for gaining further insight into the finite-temperature behaviour of the
system. 
Here, we wish to clarify the
nature of degrees of freedom associated with the fluctuations of 
$\Delta_i$, and to suggest a technique which can be used to describe the 
system characterised by  different 
behaviours of the phase and amplitude fluctuations at various temperatures. 
Following some preliminary 
considerations of the available quantum mechanical states on-site 
(Sec. \ref{sec:on-site}) and
an adaptation of the known results on the Euler angle parametrisation and 
Haar measure of the SU(4) group 
(Sec. \ref{sec:SU4}), we explicitly construct the corresponding set of coherent 
states on-site and 
write down the phase-space integration measure 
(Secs. \ref{sec:SU4}--\ref{sec:coh}). 
While the published
work on the FKM mainly deals with the half-filled case, we consider the general
situation of $n<2$. Although an actual implementation of a mean-field scheme 
is relegated to a future publication, in Sec. 
\ref{sec:discu} we provide
a crude tentative estimate of the phase-fluctuations contribution to the 
specific heat. 
We believe this is a fitting illustration of the physical contents and 
experimental relevance of the present study.
   
\section{Hybridisation and the on-site Hilbert space}
\label{sec:on-site}

Generally, an electronic state $| \psi \rangle$ at a given site $i$ is 
written as
\begin{eqnarray}
| \psi \rangle =&& 
{\rm e}^{{\rm i} \gamma_c}\beta_c  c^\dagger | 0 \rangle +
 {\rm e}^{{\rm i} \gamma_c+ \phi} \beta_dd^\dagger |0 \rangle + \nonumber \\
&&+\beta_0 |0 \rangle +
{\rm e}^{{\rm i} \gamma_{cd}}\beta_{cd} c^\dagger d^\dagger |0 \rangle\,.
\label{eq:arbstate}
\end{eqnarray}
Here, the four real and positive coefficients $\beta_a$ are subject to the 
normalisation condition,
$\sum_a \beta_a^2=1$, and the phases $\gamma_c$, $\gamma_{cd}$, $\phi$ vary 
from 0 to $2 \pi$. $|0 \rangle $ is the vacuum (empty) state, 
and the site index $i$ shall be suppressed forthwith.

The coefficients in (\ref{eq:arbstate}) are related to the on-site physical 
quantities as follows:
\begin{eqnarray}
n \equiv \langle \psi | c^\dagger c + d^\dagger d | \psi \rangle &=& 
\beta_c^2+\beta_d^2 + 2 \beta_{cd}^2  \,,\label{eq:nbeta}\\
n_d \equiv \langle \psi | d^\dagger d | \psi \rangle &=& \beta_d^2 + 
\beta_{cd}^2\,,\\
n-2n_d & = & \beta_c^2 - \beta_d^2\,, \\
\Delta\equiv \langle \psi | c^\dagger d | \psi \rangle &=& 
{\rm e}^{{\rm i} \phi}\beta_c \beta_d \,.
\end{eqnarray} 
In particular, we see that the hybridisation $\Delta$ arises only if both 
singly-occupied 
components $c^\dagger |0\rangle$ and $d^\dagger |0\rangle$ are present in $| \psi \rangle$, 
and its phase $\phi$ is determined
by the relative phase of the two coefficients.

Let us now perform an SU(2) transformation of operators $c$ and $d$ according 
to
\begin{eqnarray}
c^\dagger &=& \cos \frac{\theta}{2} \tilde{c}^\dagger -  {\rm e}^{{\rm i} \phi}
 \sin \frac{\theta}{2} 
\tilde{d}^\dagger\,,
\label{eq:transSU2-1}\\
d^\dagger &=&  {\rm e}^{-{\rm i} \phi}\sin \frac{\theta}{2} 
\tilde{c}^\dagger\ + \cos \frac{\theta}{2} 
 \tilde{d}^\dagger\,, 
\label{eq:transSU2-2}
\end{eqnarray} 
where $\theta$ takes values between 0 and $\pi$ and
\begin{equation}
\cos \theta = \frac{n-2n_d}{\sqrt{(n-2n_d)^2+4|\Delta|^2}}\,.
\label{eq:cosnd}
\end{equation}
Substituting this into Eq. (\ref{eq:arbstate}), we find after simple algebra:
\begin{eqnarray}
| \psi \rangle =&&{\rm e}^{{\rm i} \gamma_c}\sqrt{n-2 \goth{n_d}}
 \tilde{c}^\dagger | 0 \rangle + 
\sqrt{1-n+\goth{n_d}} \,|0 \rangle + \nonumber\\
&&+{\rm e}^{{\rm i} \gamma_{cd}}\sqrt{\goth{n_d}}\tilde{c}^\dagger 
\tilde{d}^\dagger |0 \rangle\,.
\label{eq:statetilde}
\end{eqnarray}
Here,
\begin{equation}
\goth{n_d} \equiv \langle \psi | \tilde{d}^\dagger \tilde{d} | \psi \rangle
= \frac{1}{2} \left(n-\sqrt{(n-2n_d)^2+4|\Delta|^2} \right)
\end{equation}
is the average occupancy of the new fermion corresponding to the operator 
$\tilde{d}$, which does not hybridise with $\tilde{c}^\dagger$,
\begin{equation}
\langle \psi | \tilde{c}^\dagger \tilde{d} | \psi \rangle=0\,,\,\,\,\,
\langle \psi | \tilde{c}^\dagger \tilde{c}+\tilde{d}^\dagger \tilde{d}  | \psi \rangle=n
.
\label{eq:diagstate}
\end{equation}
Reversing the transformation given by Eqs. 
(\ref{eq:transSU2-1}--\ref{eq:cosnd}),
\begin{equation}
\tilde{c}^\dagger = \cos \frac{\theta}{2} {c}^\dagger + {\rm e}^{{\rm i} \phi}
 \sin \frac{\theta}{2} 
{d}^\dagger\,
\end{equation} 
and substituting into Eq.(\ref{eq:statetilde}), we find for the coefficients 
$\beta_a$ in Eq. (\ref{eq:arbstate}):
\begin{eqnarray}
\beta_c = \sqrt{n-2 \goth{n_d}} \cos \frac{\theta}{2} ,&&\,\,\,
\beta_d = \sqrt{n-2 \goth{n_d}} \sin \frac{\theta}{2}\,,\nonumber \\
\beta_0=\sqrt{1-n+\goth{n_d}} ,&&\,\,\,\beta_{cd}=\sqrt{\goth{n_d}}\,.
\label{eq:beta}
\end{eqnarray}
The physical variables $\Delta$ and $n_d$ are thus given by
\begin{eqnarray}
\Delta&=&\frac{1}{2}{\rm e}^{{\rm i} \phi}(n-2\goth{n_d})\sin\theta\,,
\label{eq:Deltathetaphi}\\
n_d&=&\frac{1}{2} [n-(n-2\goth{n_d})\cos\theta]\,.
\label{eq:ndthetaphi}
\end{eqnarray}
These results were obtained by  transforming the fermion operators while 
keeping the state $|\psi\rangle$ constant. Alternatively, we can start from a 
state [cf. Eq. (\ref{eq:statetilde})]
\begin{eqnarray}
| \tilde{\psi} \rangle =&&{\rm e}^{{\rm i} \gamma_c}\sqrt{n-2 \goth{n_d}}
 {c}^\dagger | 0 \rangle + 
\sqrt{1-n+\goth{n_d}} \,|0 \rangle + \nonumber\\
&&+{\rm e}^{{\rm i} \gamma_{cd}} \sqrt{\goth{n_d}}{c}^\dagger 
{d}^\dagger |0 \rangle\,,
\end{eqnarray}
and consider the transformation of this state under a substitution
\begin{eqnarray}
{c}^\dagger  &\rightarrow & \cos \frac{\theta}{2} {c}^\dagger + {\rm e}^{{\rm i} \phi}
 \sin \frac{\theta}{2} 
{d}^\dagger\,,
\label{eq:transSU2-4}\\
{d}^\dagger  &\rightarrow  & - {\rm e}^{-{\rm i} \phi}
 \sin \frac{\theta}{2} {c}^\dagger  + \cos \frac{\theta}{2} {d}^\dagger  
\,.
\label{eq:transSU2-4a}
\end{eqnarray}
By varying the values of $\theta$ and $\phi$, we will sweep the entire subset of
states $|\psi(\gamma_c,\gamma_{cd},n,\goth{n_d},\theta,\phi)\rangle$ 
corresponding 
to our fixed values of the first four parameters. These states have the form
(\ref{eq:arbstate}) with the coefficients $\beta_a$ from Eqs. 
(\ref{eq:beta}) and the values of $n_d$ and $\Delta$ given by Eqs.
(\ref{eq:Deltathetaphi}--\ref{eq:ndthetaphi}).

The entire space of on-site electronic states (\ref{eq:arbstate}) is spanned 
by  
generic SU(4) 
transformations of any given state $|\psi \rangle$. The SU(2) transformations
such as (\ref{eq:transSU2-4}--\ref{eq:transSU2-4a}) form a subgroup of the SU(4) group.

\section{The SU(4) group: parametrisation of a vector}
\label{sec:SU4}

A generic SU(4) transformation ${\cal D}$ is parametrised by fifteen Euler
angles $\alpha_a$ as\cite{Tilma2002}
\begin{eqnarray}
&&\!\!\!\!\!\!\!\!\!\!\!\!\!\!\!\!\!\!\!\!\!\!\!\!
{\cal D}={\rm e}^{{\rm i} \alpha_1 \lambda_3}
{\rm e}^{{\rm i} \alpha_2 \lambda_2}
{\rm e}^{{\rm i} \alpha_3 \lambda_3}
{\rm e}^{{\rm i} \alpha_4 \lambda_5}
{\rm e}^{{\rm i} \alpha_5 \lambda_3}
{\rm e}^{{\rm i} \alpha_6 \lambda_{10}}
{\rm e}^{{\rm i} \alpha_7 \lambda_3}
{\rm e}^{{\rm i} \alpha_8 \lambda_2}\times \nonumber \\
&&\!\!\!\!\!\!\!\!\!\!\!\!\!\!\!\!\!\!\!\!\!\!\!\!
\times {\rm e}^{{\rm i} \alpha_9 \lambda_3} 
{\rm e}^{{\rm i} \alpha_{10} \lambda_5}
{\rm e}^{{\rm i} \alpha_{11} \lambda_3}
{\rm e}^{{\rm i} \alpha_{12} \lambda_2}
{\rm e}^{{\rm i} \alpha_{13} \lambda_3}
{\rm e}^{{\rm i} \alpha_{14} \lambda_8}
{\rm e}^{{\rm i} \alpha_{15} \lambda_{15}}\,.
\label{eq:DSU4}
\end{eqnarray}
The matrices $\lambda_a$, which are given in Eq. (A1) of Ref. \cite{Tilma2002},
are the four-dimensional analogues of the Gell-Mann matrices familiar from 
the elementary particle theory. An arbitrary vector $|\psi\rangle$ in the 
four-dimensional Hilbert state can be obtained by ${\cal D}$ acting on a vector
$|\psi_0\rangle$, which we choose as
\begin{equation}
|\psi_0\rangle=\left(\begin{array}{l} 0\\ 0\\ 0\\ 1 \end{array} \right)\,,
\,\,\,\,\,|\psi\rangle ={\cal D} |\psi_0\rangle\,.
\end{equation}
The first term in Eq. (\ref{eq:DSU4}) to act on $|\psi_0\rangle$ contains a 
diagonal matrix
$\lambda_{15}$, and we readily find
\begin{equation}
{\rm e}^{{\rm i} \alpha_{15} \lambda_{15}}|\psi_0\rangle=\exp\left( -{\rm i} 
\sqrt{\frac{3}{2}}\alpha_{15} \right) |\psi_0\rangle\,. 
\end{equation}
Further, matrices $\lambda_a$ with $a= 1 \div 8$ have a 3x3 block structure, 
viz., $(\lambda_a)_{\mu, \nu}=0$ when at least one of either $\mu$ or $\nu$ equals
four. When exponentiated, this yields for ${\cal D}_{b,a}= \exp({\rm i} 
\alpha_b \lambda_a)$ a block-diagonal form: $({\cal D}_{b,a})_{4,4}=1$ and 
$({\cal D}_{b,a})_{\mu,4}=({\cal D}_{b,a})_{4,\mu}=0$ for $\mu\neq 4$. This means
that the next eight exponential factors in Eq. (\ref{eq:DSU4}) leave 
$|\psi_0\rangle$ invariant. In the remaining first six factors on the 
r.\ h.\ s. of Eq.(\ref{eq:DSU4}), the explicit 
exponentiation should be performed, facilitated by the similarity of the
corresponding $\lambda_a$ to the Pauli matrices. Ultimately, one finds
\begin{equation}
\!\!\!\!\!\!\!\!\!\!|\psi\rangle =
{\rm e}^{ -{\rm i} 
\sqrt{\frac{3}{2}}\alpha_{15}}
\left(\begin{array}{l}
{\rm e}^{{\rm i} (\alpha_1+\alpha_3+\alpha_5)} \cos \alpha_2 \cos \alpha_4 
\sin \alpha_6\\
-{\rm e}^{{\rm i} (-\alpha_1+\alpha_3+\alpha_5)} \sin \alpha_2 \cos \alpha_4 
\sin \alpha_6\\
-{\rm e}^{{\rm i} \alpha_5}  \sin \alpha_4 \sin \alpha_6\\
\cos \alpha_6
\end{array} \right)\,,
\label{eq:psialpha16}
\end{equation}
where we can drop the exponential pre-factor. Thus, an arbitrary state
$|\psi\rangle$ is parametrised by six real variables $\alpha_a$ with 
$a = 1 \div 6$, as could have been anticipated based on the discussion in the
previous section.

It is possible to perform integrations over the group space using a measure which is invariant under
the group action (Haar measure). Following Ref. \cite{Tilma2002} we write this as
\begin{eqnarray}
d \Omega=&&\frac{24}{\pi^3} \cos^3 \alpha_4 \cos \alpha_6 \sin 2 \alpha_2 \sin \alpha_4 \times
\nonumber \\
&&\times \sin^5 \alpha_6 \,\, d \alpha_1 d \alpha_2 d \alpha_3 d \alpha_4 d \alpha_5 d \alpha_6\,, 
\label{eq:Haar1}
\end{eqnarray}
where on the r.\ h.\ s. we omitted the product of $d \alpha_a$ with $a = 7 \div 15$, 
as the vector $|\psi \rangle$, Eq. (\ref{eq:psialpha16}), 
does not depend on the corresponding $\alpha_a$. 
In order to sweep the entire Hilbert space once, the remaining $\alpha_a$'s should vary in the following 
intervals:
\begin{eqnarray}
&& 0\leq \alpha_1 \leq \pi\,,\,\,\,\,\, 0 \leq \alpha_2, \alpha_4, \alpha_6 \leq \frac{\pi}{2}\,\nonumber \\
&& 0 \leq \alpha_3, \alpha_5 \leq 2 \pi
\label{eq:rangealpha}
\end{eqnarray}
[see Ref. \cite{Tilma2002}, Eq. (C7)]. Our choice of the pre-factor in Eq. (\ref{eq:Haar1})
corresponds to the net volume equal to the total number of states (four):
\begin{equation}
\int_{SU(4)} d \Omega=4\,.
\end{equation}
The subscript SU(4) denotes integration over the entire range specified by the inequalities (\ref{eq:rangealpha}).

In a direct analogy to the spin-coherent states generated by SU(2) 
rotations\cite{Assa}, the states 
$|\psi(\alpha_1,\alpha_2,\alpha_3,\alpha_4,\alpha_5,\alpha_6)\rangle$, 
given by Eq. (\ref{eq:psialpha16}),
form an overcomplete basis of {\it SU(4) coherent states} in our four-dimensional Hilbert space. Indeed,
it is straightforward to verify the resolution of unity,
\begin{equation}
\int_{SU(4)} |\psi(\alpha_1,...,\alpha_6)\rangle
\langle\psi(\alpha_1,...,\alpha_6)| d \Omega=\hat{1}\,.
\end{equation}
It follows that the trace of any operator $\hat{{\cal O}}$ over the Hilbert space can be evaluated as
\begin{equation}
{\rm Tr}\hat{{\cal O}}=\int_{SU(4)} \langle\psi(\alpha_1,...,\alpha_6)| 
\hat{{\cal O}}|\psi(\alpha_1,...,\alpha_6)\rangle\, d \Omega.
\label{eq:trace}
\end{equation}

\section{Phase-space integration}
\label{sec:coh}

\begin{figure}
\includegraphics[width=.29\textwidth]{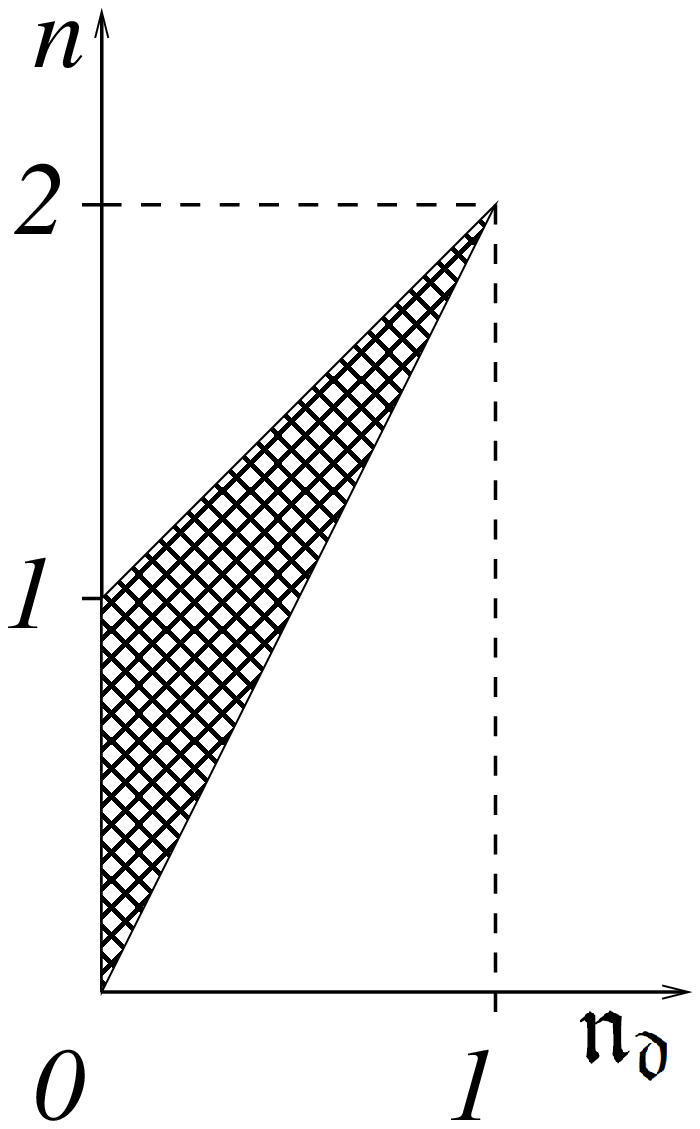}
\caption{\label{fig:ndrange} Ranges of integration over 
the $SU(4)$ group for $n$ and $\goth{n_d}$, 
Eqs. (\ref{eq:nrange}--\ref{eq:ndrange}).}  
\end{figure}

We begin with translating the mathematical results of the previous section into the language of the
electronic states on-site discussed in Sec. \ref{sec:on-site}.
Since  the vector $|\psi\rangle$ is defined up to an overall phase factor, 
we can multiply
the r.\ h.\ s. of Eq. (\ref{eq:psialpha16}) by
$-\exp( {\rm i} 
\sqrt{{3}/{2}}\alpha_{15}-{\rm i} \alpha_5)$. We then assign the 
four components of the vector, top to bottom, as corresponding to
$c^\dagger | 0 \rangle$, $d^\dagger | 0 \rangle$, $| 0 \rangle$ and 
$c^\dagger d^\dagger | 0 \rangle$. Comparing $|\psi\rangle$
to the form (\ref{eq:arbstate}) we find from Eqs. (\ref{eq:nbeta}) 
and (\ref{eq:beta}):
\[
\!\!\!\!\!\!\!\phi=-2\alpha_1+\pi\,,\,\,\,\,
\gamma_c=\alpha_1+\alpha_3+\pi\,,\,\,\,\gamma_{cd}=-\alpha_5+\pi
\]
and, with the help of Eqs. (\ref{eq:nbeta}) and (\ref{eq:beta}),
\begin{eqnarray}
\theta &=& 2 \alpha_2\,,\,\,\,\goth{n_d}=\cos^2 \alpha_6\,, \nonumber \\
n&=&\sin^2 \alpha_6 \cos^2 \alpha_4+ 2 \cos^2 \alpha_6\,.
\end{eqnarray}
Evaluating the Jacobians, $|\partial(\theta,\phi,\gamma_c,\gamma_{cd})/
\partial(\alpha_1,\alpha_2,\alpha_3,\alpha_5)|=4$ and
\begin{equation}
\frac{\partial(n, \goth{n_d})}{\partial(\alpha_4,\alpha_6)}=
\sin 2 \alpha_6 \sin 2 \alpha_4 \sin^2
\alpha_6\, 
\end{equation}
we find from Eq. (\ref{eq:Haar1}),
\begin{equation}
d\Omega=\frac{3}{2\pi^3}(n-2 \goth{n_d}) \sin \theta \, d\gamma_c 
d \gamma_{cd} dn d \goth{n_d} d \theta d \phi\,,
\label{eq:Haar2}
\end{equation}
where according to Eq. (\ref{eq:Deltathetaphi}), $(n-2 \goth{n_d}) 
\sin \theta = 2 |\Delta|$. Traces of  operators can
thus be evaluated using Eq. (\ref{eq:trace}) with 
$|\psi\rangle=|\psi(\gamma_c ,\gamma_{cd} ,n,\goth{n_d}, \theta, \phi)\rangle$ 
given by Eqs. (\ref{eq:arbstate}) and (\ref{eq:beta}).  
The integration ranges are
\begin{eqnarray}
&&0 \leq \phi, \gamma_c, \gamma_{cd} \leq 2 \pi\,,\,\,\,0 \leq \theta \leq \pi\,, \label{eq:phirange}\\
&& 0 \leq n \leq 2\,, \label{eq:nrange}\\
&&\left\{\begin{array}{l} 0 \leq \goth{n_d} \leq n/2 \,\,{\rm for} \,\, n \leq 1\,, \\
~ \\
n-1 \leq \goth{n_d} \leq n/2 \,\, {\rm for}\,\, n > 1
\end{array} \right. \label{eq:ndrange}
\end{eqnarray}
(see Fig. \ref{fig:ndrange}). The dependence of the on-site physical quantities $\Delta$ and $n_d$ on the integration
variables is given by Eqs. (\ref{eq:Deltathetaphi}--\ref{eq:ndthetaphi}).

It is instructive to calculate the number of states on-site available for a 
fixed value of $n=n_0$:
\begin{equation}
\!\!\!\!\!\!\!\!\!\!\!\!
{\cal N}(n_0)=\int_{SU(4)} \!\!\!\!\!\!\! \delta(n-n_0) d \Omega = \left\{\begin{array}{l} 6 n_0^2  \,\, {\rm for} \,\,  0 \leq n_0 \leq 1\,, \\
~\\
6(2-n_0)^2 \,\,  {\rm for} \,\,  1 < n_0 \leq 2 \end{array} \right.
\!\!\!\!\!\!\!\!\!\!\!
\end{equation}
with $\int_0^2{\cal N}(n_0) dn_0=4$, the net number of available states.

The on-site electron states in the presence of itinerant electrons are 
not pure in the quantum-mechanical sense, and should be described with a density
matrix\footnote{Traces of operators on-site, Eq. (\ref{eq:trace}), can still 
be calculated using the pure states, Eq. (\ref{eq:arbstate}). 
The integration measure actually depends on the choice of metric in the 
space of density matrices,
as will be discussed elsewhere. However, Eq. (\ref{eq:Haar2a}) is 
adequate for our
present purposes. }. If the state 
of the entire system corresponds to a fixed integer
number of electrons 
(which is a possible choice, due to overall particle conservation by the 
Hamiltonian), this on-site
density matrix will be diagonal in the number of particles on-site. In other
words, there will be no off-diagonal elements involving either $|0\rangle$ or
$c^\dagger d^\dagger|0\rangle$, unlike in a factorizable density matrix built out
of the pure states (\ref{eq:arbstate}). In order to restrict the trace in
Eq. (\ref{eq:trace}) to this subset, one should replace the generic 
operator ${\cal O}$
with $\sum_{n=0}^2 {\cal P}_n {\cal O} {\cal P}_n$, where ${\cal P}_n$ is a 
projection onto a subspace with a given value of $n$ (thus ${\cal P}_1$ projects
onto the subspace of linear combinations of $c^\dagger |0\rangle$ and
$d^\dagger|0\rangle$). Obviously, operators corresponding to the physical 
observables on-site already have this structure. Either way, the integrand
in Eq. (\ref{eq:trace}) will be independent of the phases $\gamma_c$ and 
$\gamma_{cd}$ in Eq. (\ref{eq:arbstate}), and the corresponding integration
[along with the pre-factor $1/(2\pi)^2$] can be dropped\footnote{On the other 
hand, in a {\it superconducting} state at  $U<0$ the quantity  
$\gamma_{cd}$  
acquires a physical meaning of the phase of the order parameter. This reflects 
the fact that the BCS wave function does not correspond to a fixed
number of particles.}. Thus, we finally 
arrive at  
\begin{equation}
d\Omega=\frac{6}{\pi}(n-2 \goth{n_d}) \sin \theta \, 
dn d \goth{n_d} d \theta d \phi\,,
\label{eq:Haar2a}
\end{equation}
where the integration region for the four variables is still given by Eqs.
(\ref{eq:phirange}--\ref{eq:ndrange}).

\section{Prolegomena to the mean field theory}
\label{sec:discu}

The formalism developed in the previous sections provides necessary 
information about
the structure of the phase space of the on-site 
variables $\Delta_i$ and $n_{d,i}$. This enables constructing a single-site 
mean field description for the extended FKM and related models. While postponing
a truly self-consistent calculation to a future publication, we will now 
briefly discuss the appreciated 
results at a rather qualitative level. We will make the following simplifying
assumptions:

\noindent
(i) The system can be described in terms of a single-site energy $E$, 
which depends on the fluctuating values of local parameters $n$,$\goth{n_d}$,
$\theta$ and $\phi$. Here, we again omit the subscript corresponding to the 
chosen site, which should be viewed as embedded into a virtual crystal 
characterised by the average values of these 
parameters. 

\noindent
(ii) Fluctuations of both $n$ and $\goth{n_d}$ are negligible, 
and their respective average values are temperature-independent\footnote{This 
assumption
will be addressed and perhaps 
modified in the course of the 
forthcoming proper treatment. Presently, we expect it  to be adequate for our 
purposes.}. This leaves two parameters $\theta$ and $\phi$,
and the integration measure is that of the SU(2) subgroup:
\begin{equation}
d\Omega^\prime=(n-2 \goth{n_d}) \sin \theta d \theta d \phi\,,
\label{eq:Haar3}
\end{equation}  
where we omitted the unknown (and unimportant) numerical pre-factor.

\noindent (iii) On-site parameters $|\Delta|$ and $\phi$ which enter 
the single-site energy $E$ 
can be treated 
as classical variables. We expect this to be qualitatively correct 
when thermal fluctuations are sufficiently strong.
At very low temperatures, on the other hand, any single-site mean-field 
approach would be inadequate.

\noindent(iv) The minimum of energy $E$ is attained at $|\Delta|=\Delta_0(T)$ 
and (in the ordered phase at $T<T_c$) at $\phi=0$.
We assume that  $\Delta_0$ equals zero above the crossover temperature 
$T_\Delta$ (where $T_\Delta \gg T_c$, see Sec. \ref{sec:intro})
 and   below $T_\Delta$ shows  typical behaviour of a
solution to a BCS-like gap equation:
\begin{equation}
\Delta_0(T)=\Delta_0(0)\sqrt{(T_\Delta-T)/T_\Delta}\,,\,\,\,T<T_\Delta\,.
\label{eq:Delta0}
\end{equation}
Quantitatively this assumption overestimates the steepness of the crossover, 
as it can be argued that $\Delta_0(T)$ never
vanishes\footnote{The gap equation for $\Delta$ holds in the uniform case 
and does not describe single-site fluctuations. Here it is referred to
for simplicity, as a crude initial approximation.}. We write in a 
Ginzburg--Landau fashion,
\begin{eqnarray}
&&E(|\Delta|,\phi)=
2A \frac{T-T_\Delta}{T_\Delta}[\Delta_0(0)]^2|\Delta|^2 + A |\Delta|^4 -
\nonumber \\
&&-B(T) \cos{\phi}+\frac{1}{2}B(T)\langle\!\langle \cos{\phi} \rangle\! \rangle\,,
\label{eq:dE}
\end{eqnarray}
omitting  terms which do not depend on $\Delta$ and $\phi$.
The coefficient $A$ does not depend on temperature, whereas the molecular field
$B(T)$ vanishes at $T>T_c$, resulting in a second-order
phase transition (loss of the long-range order of the phases $\phi_i$) at $T_c$:
\begin{equation}
B(T)=B(0)\sqrt{(T_c-T)/T_c}\,,\,\,\,T<T_c\,\,.
\end{equation}
Physically, $B(T)$ originates from the last term in Eq. (\ref{eq:FKM}).
The last term in Eq. (\ref{eq:dE}) offsets the usual mean-field energy 
double-counting, with
\begin{equation}
\!\!\!\!\!\!\!\!\!\!\!\! \langle\!\langle \cos{\phi} \rangle\! 
\rangle \equiv
\frac{1}{Z}\int \cos{\phi}
{\rm e}^{-E(|\Delta|,\phi)/T}d\Omega^\prime= \frac{I_1[B(T)/T]}{I_0[B(T)/T]}
\end{equation} 
at $T<T_c$. Here, $I_n$ are the imaginary argument Bessel functions, and $Z$
is the partition function,
\begin{eqnarray}
Z\equiv \int {\rm e}^{-E(|\Delta|,\phi)/T}d\Omega^\prime\,. 
\label{eq:partfunc}
\end{eqnarray} 
The discontinuity of specific heat $C$ at $T_c$ is obtained as
\begin{equation}
C(T_c-0)-C(T_c+0)=\frac{1}{32} \left[B(0)/T_c\right]^4\,.
\end{equation}

Formally, there are two distinct values of the angle $\theta$, corresponding
to $|\Delta|=\Delta_0(T)$  [see Eq.(\ref{eq:Deltathetaphi})] . 
At the level of our discussion here, this 
appears to be due to simplifications we made in writing Eq. (\ref{eq:dE}). 
We note, 
however, that this corresponds to the fact that the on-site density matrix
can be parametrised by {\it two} vector (``pure-state'') components, 
with the two respective values of  $\theta$ adding up to 
$\pi$ (i.e., same value of $\sin \theta$).  Since both components  show similar 
behaviour with temperature, we
will follow only one of these.
Assuming that the temperature is not too high, we may restrict 
integration over $\theta$ in Eq. (\ref{eq:partfunc}) to  $0<\theta<\pi/2$,
thereby choosing the $\theta<\pi/2$ component.
In the phase-disordered region between $T_c$ and $T_\Delta$, the energy 
$E(|\Delta|,\phi)$ can be expanded about its minimum, $|\Delta|=\Delta_0(T)$,
yielding 
\begin{equation}
\!\!\!\!\!\!\!\!\!\!\!\!\!
Z\approx 4\left\{\frac{\pi^3 T}{A[(n-2\goth{n_d})^2-(2\Delta_0(T))^2]}\right\}^{1/2}
{\rm e}^{A (\Delta_0(T))^4/T}\,,
\label{eq:Zsaddle}
\end{equation} 
provided that $T_\Delta-T$ is not too small, $(T_\Delta-T)^2 \gg TT_\Delta^2/4A$. We 
find\footnote{In addition,
there exist contributions from other degrees of freedom, such as electron-hole 
excitations and phonons, which were not included in this estimate.}
\begin{eqnarray}
\!\!\!\!\!\!\!\!\!\!\!\!\!
C\!\approx&&
\!\!\!\!\!\!\!\!\!\!\!\!
\frac{1}{2}+2 A[\Delta_0(0)]^4\frac{T}{T_\Delta^2}-\frac{T}{T_\Delta}
\frac{4[\Delta_0(0)]^2}{(n-2\goth{n_d})^2
-4[\Delta_0(T)]^2}+ 
\nonumber \\
\!\!\!\!\!\!\!\!\!\!\!\!
&&+\frac{T^2}{T_\Delta^2}\frac{8[\Delta_0(0)]^4}{\{(n-2\goth{n_d})^2
-4[\Delta_0(T)]^2\}^{2}}\,. \label{eq:C}
\end{eqnarray}
The value of $C$ increases superlinearly with temperature, and
the coefficient in the $T^2$ term decreases smoothly as the temperature 
increases towards $T_\Delta$.   

We wish to emphasise the role of
the phase degree of freedom even in the disordered state above $T_c$. Indeed, 
considering $\phi$ as a fictitious variable would lead to a substitution of 
the integration measure $d \Omega^\prime$, Eq. (\ref{eq:Haar3}), with merely 
$d\theta$ [replacing the integration over SU(2) subgroup with the U(1) one].
Clearly, the corresponding partition function $\tilde{Z}$ in this 
temperature range is given by $Z$ of 
Eq. (\ref{eq:Zsaddle}) divided by the value of 
$2 \pi (n-2 \goth{n_d}) \sin \theta$ at the energy minimum, 
i.e., by $4\pi \Delta_0(T)$. The result for the
specific heat would be
\begin{equation}
\tilde{C}\approx C+\frac{T}{2}\frac{2 T_\Delta-T}{(T_\Delta-T)^2}.
\end{equation} 
This difference is due to the larger relative phase space volume at small 
$\Delta$ [which otherwise is suppressed by the weight $2|\Delta|$ entering the
SU(2) integration measure, Eq. (\ref{eq:Haar3})]. This becomes important
as the temperatures increase toward $T_\Delta$ and the energy $E(|\Delta|)$ , 
Eq. (\ref{eq:dE}), softens at $\Delta=0$. Thus taking phase fluctuations 
into account results in a 
slower decrease of the average value of $|\Delta|$, and reduces the values 
of specific heat.

\begin{figure}
\includegraphics[width=.49\textwidth]{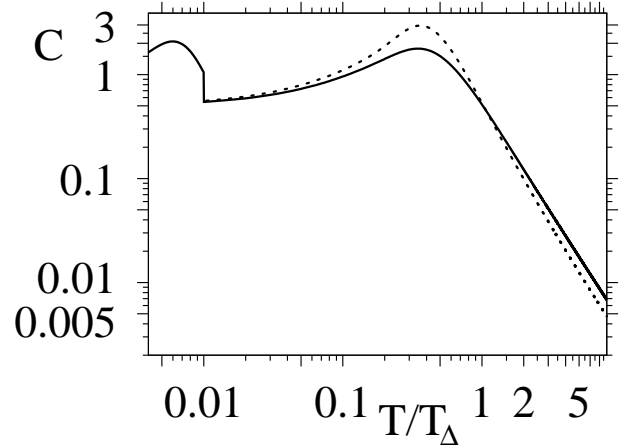}
\caption{\label{fig:C} Typical temperature dependence of specific heat $C$
(solid line). The dotted line  (plotted for $T>T_c$) represents $\tilde{C}$, 
which does not include the 
contribution of phase fluctuations to the partition function. Parameter values
are $n-2\goth{n_d}=0.5$, $\Delta_0(0)=0.125$, $B(0)=0.02$, $T_c/T_\Delta=0.01$,
and $A/T_\Delta=10^4$.} 
\end{figure}

\begin{figure}
\includegraphics[width=.49\textwidth]{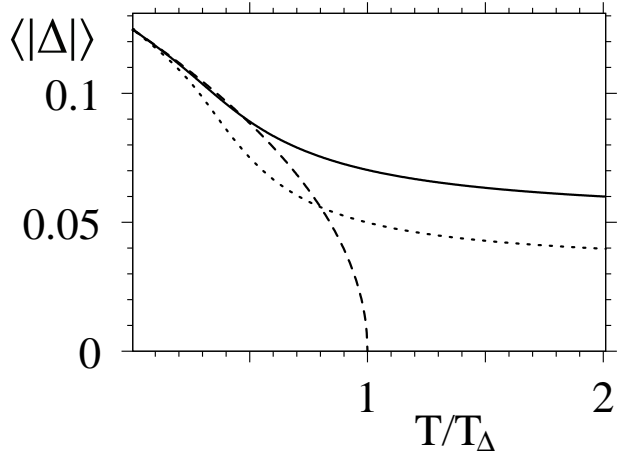}
\caption{\label{fig:Delta} Thermal average of $|\Delta|$ (solid line) for the 
values of parameters used in Fig. \ref{fig:C}. Dotted line shows the result 
obtained by neglecting phase fluctuations, whereas the dashed line corresponds 
to $\Delta_0(T)$, Eq. (\ref{eq:Delta0}).}  
\end{figure}

Typical numerical results for $C$ (solid line) and $\tilde{C}$ are shown in 
Fig. \ref{fig:C}, whereas the corresponding average values of $|\Delta|$ are
plotted in Fig. \ref{fig:Delta}.
As mentioned above, we cannot expect to obtain a faithful description of the $T\rightarrow 0$
case, which explains the finite value of $C$ found in this limit.
Following a negative jump at $T=T_c$, $C$ begins to increase as dictated by
the second term in Eq. (\ref{eq:C}). This increase [which at higher $T$ is 
suppressed
by terms omitted in Eq. (\ref{eq:C})] becomes less pronounced and eventually disappears if the coefficient
$A$ in Eq. (\ref{eq:dE}) is decreased. As noted above, omitting phase
fluctuations yields to a much stronger increase in $\tilde{C}$, with the
ratio $\tilde{C}/C$ approaching 2 at the peak value (note the log scale
in Fig. \ref{fig:C}).

Above the crossover temperature 
$T_\Delta$, taking phase 
fluctuations into account reduces the available phase space volume near
$|\Delta|=0$. This leads to stronger fluctuations of $|\Delta|$
(and accordingly yields a larger average value of $|\Delta|$) and to 
an increased specific heat. In fact, if
the parameter values allow for the regime where
\begin{equation}
(n-2\goth{n_d})^2 \gg [\Delta_0(0)]^2 \frac{T-T_\Delta}{T_\Delta} \gg 
\sqrt{\frac{T}{A}}\,,
\end{equation}
we find
\begin{equation}
C \approx \frac{T_\Delta^2}{(T-T_\Delta)^2} \approx 2 \tilde{C}\,.
\end{equation}

These results, while tentative, highlight the importance of correctly
taking into account the available phase space volume in a disordered excitonic 
insulator above $T_c$. 
While this preliminary discussion was limited to the
extended Falicov--Kimball model, we expect similar physics to play a role
in related systems, including Kondo lattices.

\section{Acknowledgements}
The author takes pleasure in thanking A. G. Abanov, R. Berkovits, A. Frydman, 
A. V. Kazarnovski-Krol, and M. Khodas
for discussions. This work was supported by the Israeli 
Absorption Ministry.


\end{document}